\documentclass{aa}

\usepackage{graphicx}
\usepackage{txfonts}
\usepackage{siunitx}
\usepackage{hyperref}

\begin{document}

\title{First detection of ozone in the mid-infrared at Mars: implications for methane detection}
\author{K. S. Olsen\inst{1,2} 
\and F. Lef\`evre\inst{2} 
\and F. Montmessin\inst{2}
\and A. Trokhimovskiy\inst{3}
\and L. Baggio\inst{2}
\and A. Fedorova\inst{3}
\and J. Alday\inst{1}
\and A. Lomakin\inst{3,4}
\and D. A. Belyaev\inst{3}
\and A. Patrakeev\inst{3}
\and A. Shakun\inst{3}
\and O. Korablev\inst{3}}
\institute{Department of Physics, University of Oxford, Oxford, UK
\\\email{Kevin.Olsen@physics.ox.ac.uk}
\and Laboratoire Atmosph\`eres, Milieux, Observations Spatiales (LATMOS/CNRS), Paris, France 
\and Space Research Institute (IKI), Moscow, Russia
\and Moscow Institute of Physics and Technology, Moscow, Russian Federation}
\date{Received April 8, 2020}

\abstract{}
{The ExoMars Trace Gas Orbiter (TGO) was sent to Mars in March 2016 to search for trace gases 
diagnostic of active geological or biogenic processes.}
{We report the first observation of the spectral features of Martian ozone (O$_3$) in the 
mid-infrared range using the Atmospheric Chemistry Suite (ACS) Mid-InfaRed (MIR) channel, 
a cross-dispersion spectrometer operating in solar occultation mode with the finest spectral 
resolution of any remote sensing mission to Mars.} {Observations of ozone were made at high 
northern latitudes ($>65^\circ$N) prior to the onset of the 2018 global dust storm 
($\mathrm{L}_\mathrm{s}=163$--193$^\circ$). During this fast transition phase between 
summer and winter ozone distribution, the O$_3$ volume mixing ratio observed is 100--200~ppbv near 
20~km. These amounts are consistent with past observations made at the edge of the southern 
polar vortex in the ultraviolet range. The observed spectral signature of ozone at 
3000--3060~cm$^{-1}$ directly overlaps with the spectral range of the methane (CH$_4$) $\nu_3$ 
vibration-rotation band, and it, along with a newly discovered CO$_2$ band in the same region,
may interfere with measurements of methane abundance.}
{}

\maketitle

\section{Introduction}
Ozone (O$_3$) on Mars was first observed by the Ultraviolet Spectrometers on Mariner 7 and 9 
\citep{Barth73,Barth71}, which showed large variability and established seasonal trends. Since 
then, O$_3$ has been observed by ground-based campaigns and spacecraft missions, but largely 
using absorption and emission features in the ultraviolet spectral range 
\citep[e.g.][]{Clancy16,Perrier06} or the thermal spectral range using the 9.7~\si{\um} band 
\citep{Espenak91,Fast06}. Here, we report the first observations of ozone absorption in the 
mid-infrared spectral region, between 3015 and 3050~cm$^{-1}$, using the mid-infrared channel 
of the Atmospheric Chemistry Suite (ACS MIR) onboard the ExoMars Trace Gas Orbiter (TGO). This spectral 
region is shared by the $\nu_3$ vibration-rotation band of methane (CH$_4$), as well as a 
newly discovered transition of CO$_2$ \citep{Trokh20}. The ability to simultaneously 
resolve these species has an impact on current and past attempts to measure the abundance of 
methane in the atmosphere of Mars.

ACS MIR is a novel cross-dispersion spectrometer making solar occultation observations of the 
limb of the Martian atmosphere. It has the finest spectral resolution of any Martian remote-sensing 
instrument to date, and the solar occultation technique provides a high signal-to-noise ratio (S/N), 
and strong sensitivity to the vertical structure of the atmosphere. These characteristics were 
instrumental in observing, for the first time, the mid-infrared 003$\leftarrow$000 transitions of 
O$_3$ in the atmosphere of Mars.

Here we present solar occultation observations made in Mars year (MY) 34 by ACS MIR between solar 
longitude ($\mathrm{L}_\mathrm{s}$) 163--193$^\circ$ and north of 60$^\circ$N (May--June 2018). 
In this region and time period, 
corresponding to the northern autumn equinox, we were able to observe significant amounts of ozone  
in the mid-infrared at altitudes below 30~km. In the following section, we describe the TGO 
mission, ACS instrument, and spectral fitting method. In Section~\ref{sec:obs} we present our 
observations, analysis, and comparison to model results and the literature. Section~\ref{sec:ch4}  
discusses the implications that this observation has for CH$_4$, which is sought in the same wavenumber 
range of the infrared.

\begin{figure*}[h!]
   \includegraphics[width=18cm]{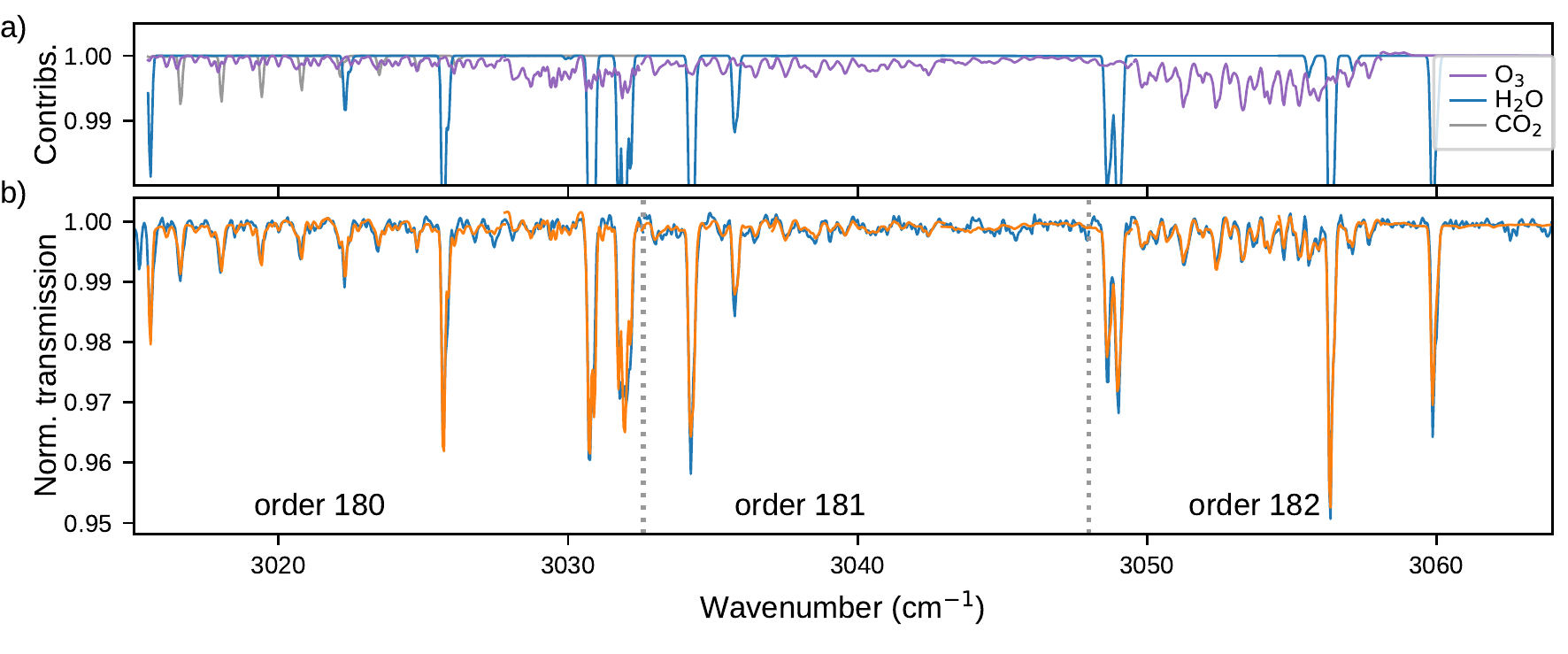}
   \caption{Spectra recorded by ACS MIR at 5.5~km using secondary grating position 12 during occultation
   2476 N1 on $\mathrm{L}_\mathrm{s}=192.7^\circ$: a) contributions from O$_3$, H$_2$O, and CO$_2$ to the 
   best-fit for orders 180--182; b) data and best-fits for orders 180--182.}
   \label{fig:f1}
\end{figure*}

\subsection{Ozone chemistry}
\label{sec:oz}
The stability of the CO$_2$ atmosphere on Mars depends on the abundances of odd hydrogen (H, OH, HO$_2$), which 
is a product of H$_2$O photolysis. The primary source of O$_3$ is a three-body reaction between O$_2$ and O, 
while the primary loss mechanism is the inverse reaction via photolysis. Such a cycle is neutral in terms 
of the quantity of odd oxygen remaining because O is converted to O$_3$ and vice versa. The main net-loss 
pathway of the odd oxygen family $\text{O}_\text{x}=\text{O}+\text{O}_3$ is the reaction with HO$_2$ 
($\text{HO}_2+\text{O}\rightarrow\text{OH}+\text{O}_2$), which reduces O$_\text{x}$. Ozone is 
therefore a valuable tracer of the odd hydrogen chemistry that stabilises the chemical composition of 
Mars' atmosphere, as H, OH, or HO$_2$ have never been directly measured. Since odd hydrogen is 
primarily produced by H$_2$O photolysis, ozone is expected to be anti-correlated with water vapour 
\citep[e.g.][and references therein]{Clancy96,Haberle17_chem,Perrier06}.

The ozone profiles described here were all obtained at high northern latitudes around autumn equinox 
($\mathrm{L}_\mathrm{s}=160$--190$^\circ$). At these latitudes and during this period, ultraviolet 
measurements performed 
in nadir geometry show ozone columns that rapidly increase with time \citep{Clancy16,Perrier06}. This 
dramatic rise in ozone occurs in conjunction with the buildup of the winter polar vortex and the quick 
decline in water vapour and O$_3$-destroying odd hydrogen. Observations of the ozone profile are sparse 
in the literature and do not cover the latitudes and local times sampled here. Indeed, almost all 
published profiles of ozone at high latitudes were measured later in the season, in the polar night 
\citep{Groller18,Montmessin13,Perrier06}. The only exceptions are the four polar ozone profiles 
measured by solar occultation in the ultraviolet range by \citep{Piccialli19}, but those were 
obtained in the southern hemisphere and at the edge of a fully developed polar vortex 
($\mathrm{L}_\mathrm{s}56$--68$^\circ$). 
Our knowledge of the vertical distribution of ozone is therefore still very limited, in particular 
in twilight conditions. However, all previous studies show that the largest ozone densities on Mars 
are always found in the polar vortices and that the polar ozone layer is usually located low in the 
atmosphere, typically between the surface and 20--25~km altitude.

\section{Methods}
\label{sec:meth}
The TGO was launched in 2016, began its nominal science phase in April 2018, and has completed 2 years of 
observations at the time of publication. The primary scientific objectives of TGO are to detect any 
trace gases diagnostic of active geologic or biogenic activity, characterise and attempt to locate
the possible sources of such trace gases, and characterise the water cycle on Mars \citep{Vago15}. 
To achieve these goals, the TGO carries two suites of multi-channel spectrometers: ACS 
\citep{Korablev18}, and the Nadir and Occultation for Mars Discovery (NOMAD) \citep{Vandaele18}. 
Both spectrometer suites have three channels and are capable of making 
observations in nadir, limb, and solar occultation viewing geometries.

The MIR channel of ACS is a cross-dispersion spectrometer consisting of a primary echelle grating, and a 
secondary grating used to separate diffraction orders \citep{Korablev18}. The secondary grating rotates 
through several positions to access different simultaneous spectral ranges. This work uses position 12, 
which covers the spectral range 2850--3250~cm$^{-1}$ that contains the main CH$_4$ absorption band. This 
channel has the finest spectral resolution (0.043--0.047~cm$^{-1}$) of any atmospheric remote sensing 
Mars mission and operates solely in solar occultation mode, benefiting from high signal strength, long 
optical path length, and the ability to directly probe vertical structure.

Calibration for these observations was performed at Russia's Space Research Institute (IKI) and 
involves performing several corrections to the data before subtracting a dark signal from observations 
taken over a series of tangent altitudes and a solar reference measured above the top of the atmosphere. 
Transmission spectra are obtained by dividing the corrected absorption spectra by the solar reference. 
Prior to transmission calculation, corrections applied to the data include removing dead or saturated 
pixels, accounting for subpixel drifts over time, and an ortho-rectification procedure needed to 
extract one-dimensional spectra from the two-dimensional detector array. See
\citet{Fedorova20}, \citet{Olsen20}, \citet{Trokh20} for more details.

The instantaneous field of view (IFOV) of the instrument optics observing the edge of the solar disc 
is on the order of 1--4~km when projected at the Martian limb. The Sun edge is imaged onto the detector 
array over approximately 15 rows, and from each row, a spectrum can be extracted. We use the intensity 
curve, which is a column of the detector array image and covers the vertical fields of view of each 
diffraction order (see \citet{Olsen20}), to identify a spectrum near the slit edge, which is closest 
to the centre of the solar disc.

\begin{figure*}[h!]
   \includegraphics[width=18cm]{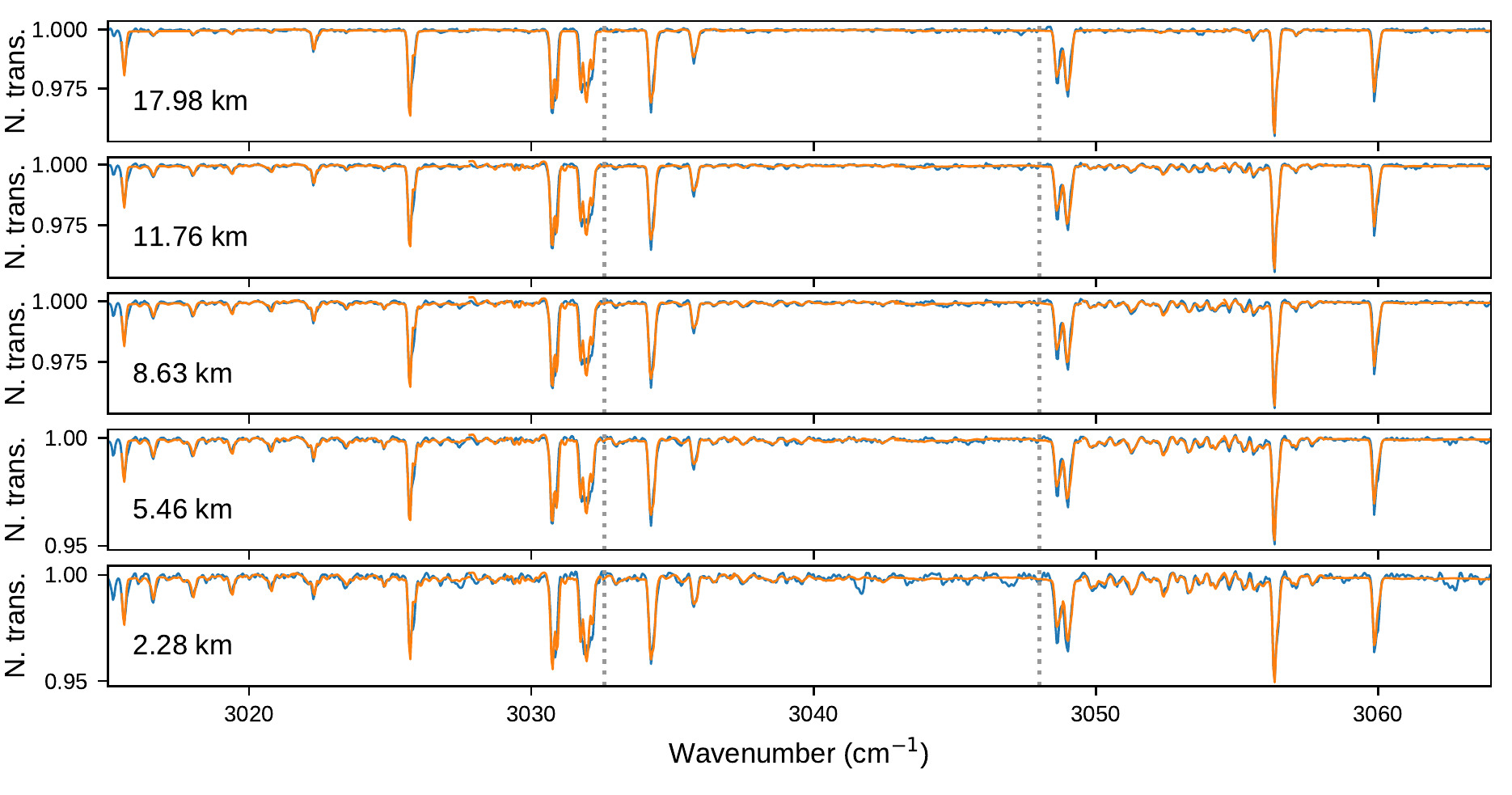}
   \caption{Same as Fig.~\ref{fig:f1}b, except at a range of altitudes.}
   \label{fig:f2}
\end{figure*}

Wavenumber calibration of each diffraction order is performed in two steps. An initial guess of wavenumber 
positions for the solar reference spectrum is compared to the solar spectrum measured by the Atmospheric 
Chemistry Experiment Fourier Transform Spectrometer (ACE-FTS) \citep{ACEsolar} in order to obtain a calibrated 
wavenumber vector. This is then refined using an appropriately clear transmission observation (avoiding 
signal attenuation due to aerosols) using transmission lines from CO$_2$, CO, or H$_2$O, if they are 
available and of sufficient strength.

Spectral fitting is performed by the Gas Fitting software suite (GFIT or GGG) maintained by NASA's 
Jet Propulsion Laboratory
\citep[e.g.][]{ATMOSret,Wunch11}. Over a given fitting window, volume absorption coefficients are 
computed for each gas and a spectrum is computed line-by-line. Non-linear Levenberg-Marquardt minimisation 
is done to determine a best-fit spectrum by modifying volume mixing ratio (VMR) scaling factors for a set 
of target gases. A set of estimated slant column abundances for all observed tangent altitudes is inverted 
with calculated slant column paths traced through the atmosphere using a linear equation solver to obtain 
a retrieved VMR vertical profile. Volume absorption coefficients are computed using the HITRAN 2016 line list \citep{HITRAN16,Olsen19}, 
supplemented by H$_2$O broadening parameters for a CO$_2$-rich atmosphere \citep{Devi17c,Gamache16}.

The width of an instrument line shape is wavenumber dependent. Therefore, to obtain more accurate fits 
we use narrow fitting windows. The SNR of each spectrum, corresponding to a diffraction order, is also 
highest in its centre due to the blaze function of the echelle grating. To avoid errors in spectral 
calibration, we avoid working toward the edges of the spectra and cover the centre of each order with 
two or three windows that are 5--6~cm$^{-1}$ in width (a diffraction order in position 12 has a width of 19--21~cm$^{-1}$). There remains a curvature in the baseline of the spectra. We use normalised spectra here and use 
an alpha shape, which finds the geometric area containing a spectrum, to estimate the baseline shape \citep{Xu19}.

Accurate knowledge of temperature and pressure are critical for computing the number density of the 
atmosphere and performing accurate trace gas retrievals. In this work, temperature and pressure have been 
retrieved from observations of CO$_2$ lines using coincident measurement made with the ACS near-infrared 
channel \citep{Fedorova20}.

\begin{figure*}
   \includegraphics[width=18cm]{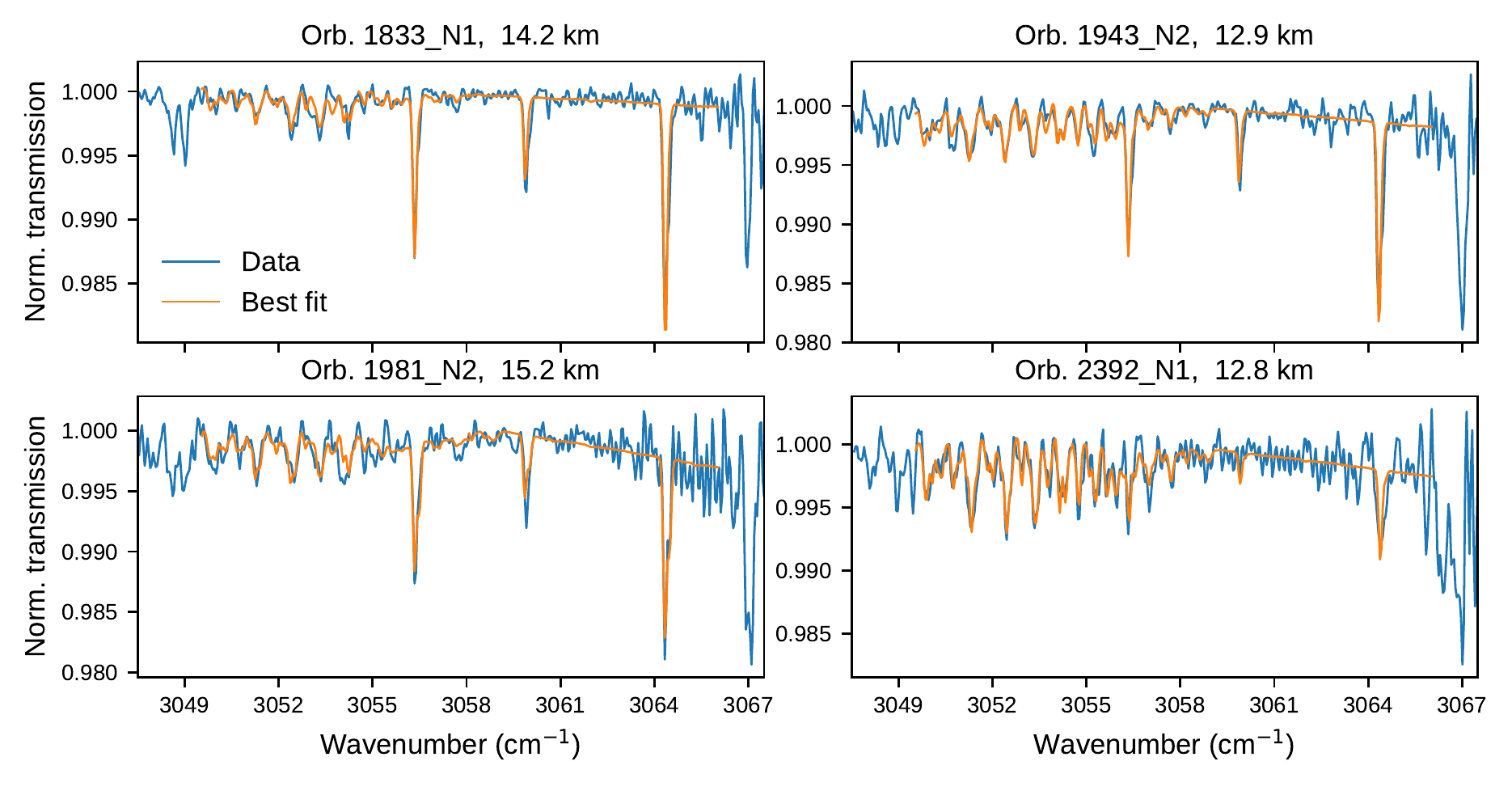}
   \caption{Measured spectra and best-fits for order 182 for four occultations recorded between 
   $\mathrm{L}_\mathrm{s}=160$--200$^\circ$ and north of 65$^\circ$N.}
   \label{fig:f3}
\end{figure*}

\begin{figure}
\centering
   \includegraphics[width=8cm]{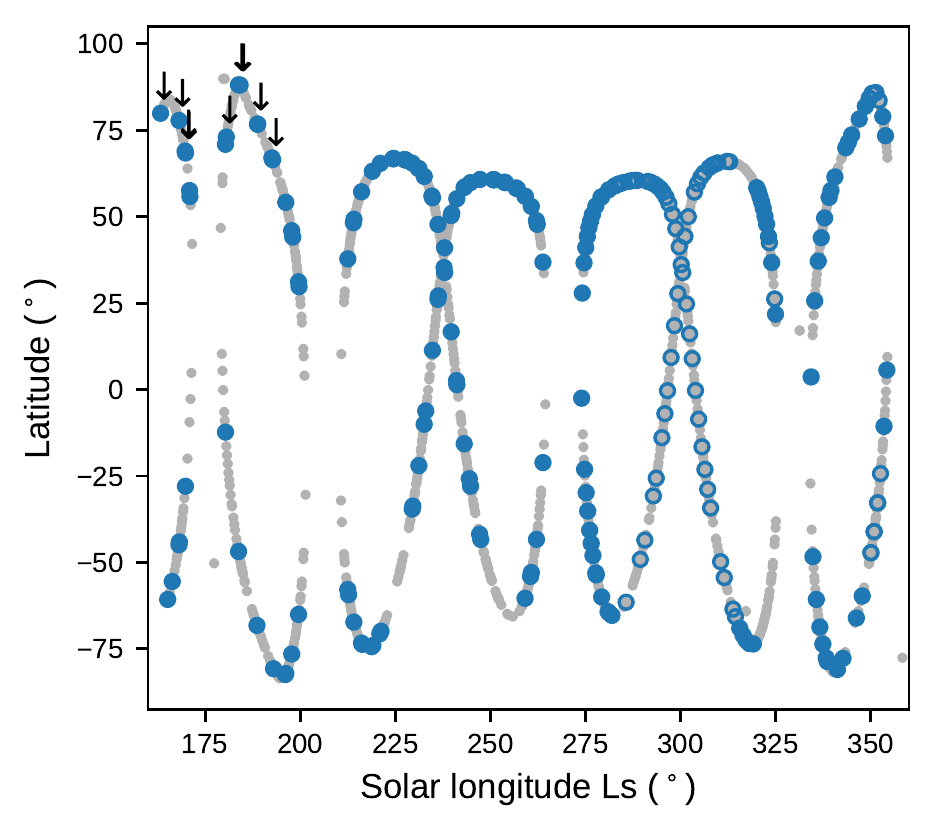}
   \caption{Latitudes of ACS MIR occultation tangent points as a function of $\mathrm{L}_\mathrm{s}$. 
   Dots in grey are 
   ACS MIR occultations using grating positions other than 12. Filled blue circles are full-frame 
   observations, and empty blue circles are partial-frame observations. Occultations exhibiting strong 
   O$_3$ absorption features are indicated with arrows.}
   \label{fig:f4}
\end{figure}

\section{Observations and Results}
\label{sec:obs}
We first discovered the transmission signature of O$_3$ in ACS MIR data during orbit 2476 (occultation N1), 
which was recorded on June 13 2018, or $\mathrm{L}_\mathrm{s}=192.7^\circ$. The local time was 17:16, 
and the latitude and longitude of the tangent point were 66.5$^\circ$N, 13.5$^\circ$E. This was shortly 
after the onset of the global dust storm of MY 34, which began around $\mathrm{L}_\mathrm{s}=190^\circ$ 
\citep{Montabone20}, but at such a high latitude that its impact was not yet felt and observations with
suitable transmittance were made down to 
2.3~km above the aeroid. The signature was initially identified in order 182, covering the spectral 
range 3047.5--3067.5~cm$^{-1}$, shown in Fig.~\ref{fig:f1}c, with a computed best-fit spectrum. Once 
identified, the signature was also clearly identified in orders 180 and 181, shown in Fig.~\ref{fig:f1}b. 
The signature is very small relative to nearby H$_2$O absorption lines and is only apparent in a small 
number of observations, and at the lowest altitude levels. It is consistently visible over a range of 
observed tangent altitudes, and its evolution over altitude is also clearly distinguishable. Spectra 
from five consecutive tangent altitudes from the same occultation are shown in Fig.~\ref{fig:f2}, from the 
lowest observable tangent altitude of 2.3~km (which had a transmission level of 30\%), up to 18~km, above 
which the signature vanishes in the instrument noise (we note there was a corrupt observation near 15~km). 
The CO$_2$ lines towards the left side of order 180 in Fig.~\ref{fig:f1}a and ~\ref{fig:f1}b are not 
included in any current line lists and have been identified as a magnetic dipole absorption band of the 
628 isotopologue of CO$_2$; they are reported in a separate paper \citep{Trokh20} and the fit shown in 
Fig.~\ref{fig:f1}b uses line positions calculated in \citep{Trokh20}, and line strengths and widths 
estimated from the nearby $\nu_2+\nu_3$ band of the $^{16}$O$^{12}$C$^{18}$O isotopologue of CO$_2$. 
The spectral window covering this area was not used to estimate the O$_3$ abundance since the O$_3$ 
transitions here are relatively weak and the CO$_2$ spectroscopy is still imprecise. 

After the initial discovery, a careful search of all processed position 12 observations was undertaken. The
signature was found in several other occultations, and often at higher altitudes. Figure~\ref{fig:f3} shows 
best fits in order 182 for four other occultations at altitudes near 15~km where the signature is strong and clear.

Figure~\ref{fig:f4} indicates where O$_3$ was identified over the evolution of the ACS MIR occultation latitudes 
with time ($\mathrm{L}_\mathrm{s}$). All of these observations occurred early in the mission, and at high 
northern latitudes, greater than 60$^\circ$N and in the range $\mathrm{L}_\mathrm{s}=160$--200$^\circ$. This is 
consistent with previous studies of the climatology of Martian O$_3$ that have reported accumulation over 
the poles during fall/winter \citep{Clancy16,Perrier06}, resulting from reduced destruction pathways caused 
by low water vapour content and solar insolation. However, according to these studies, the area of 
enhanced O$_3$ should extend southward to 40$^\circ$N and cover the time period 
$\mathrm{L}_\mathrm{s}=180$--360$^\circ$. We do not find convincing signatures of O$_3$ over the northern 
extent of ACS MIR coverage in the range $\mathrm{L}_\mathrm{s}=200$--280$^\circ$ because of the onset of the global 
dust storm. Indeed, the dust storm generally sets an altitude limit on the vertical extent of solar occultation 
observations of 25--35~km, below which meaningful signal is lost. In the range $\mathrm{L}_\mathrm{s}=280$--380$^\circ$, 
data volume constraints required us to download only partial detector frames (open circles in Fig.~\ref{fig:f4}), 
which have not been fully processed.

\begin{figure*}[h!]
   \includegraphics[width=18cm]{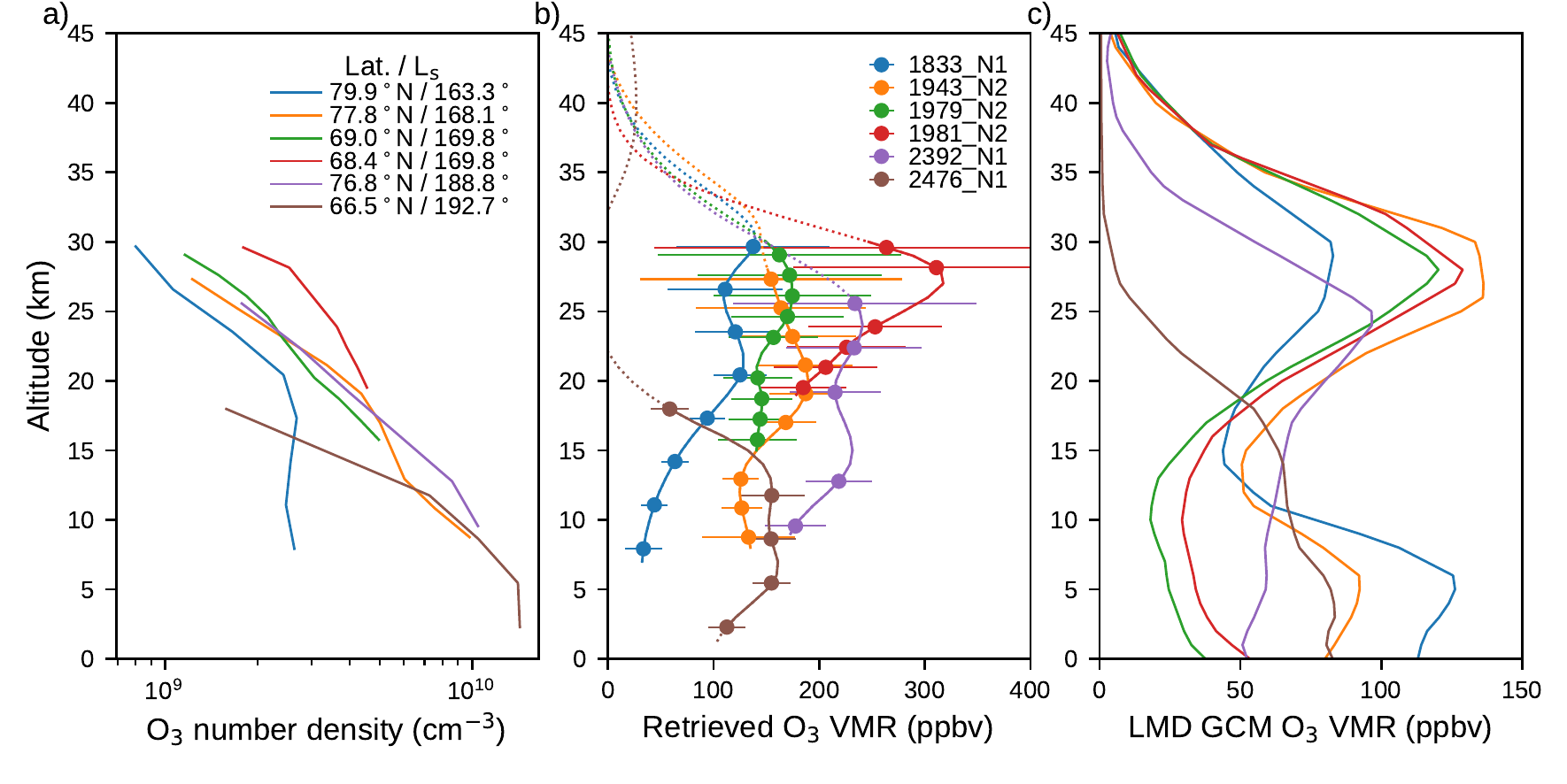}
   \caption{a) Retrieved number density or O$_3$, b) retrieved VMR vertical profiles of O$_3$, c) VMR 
   vertical profiles of O$_3$ extracted from the LMD GCM at corresponding $\mathrm{L}_\mathrm{s}$, 
   locations, and local times. 
   Colours indicating occultation number are shared between panels.}
   \label{fig:f5}
\end{figure*}

Vertical profiles of retrieved number densities and mixing ratios are shown in Fig.~\ref{fig:f5}a and b.
The lower bound of the profiles indicates where the transmission signal dropped off to less 
than 0.05. The upper altitude limit, denoted by the transition from a solid to dotted line, 
indicates where the O$_3$ absorption features begin to be hidden in background noise, causing the 
retrieval uncertainty to grow larger than the VMR. 
The uncertainties shown in Fig.~\ref{fig:f5}b are the sum of the partial derivatives computed for 
the inversion of the matrices containing the number densities and slant paths (Jacobian matrix).
In the right panel, Fig.~\ref{fig:f5}c, we also show vertical profiles generated by running the LMD General 
Circulation Model (GCM) \citep{Forget99,Lefevre04} using a dust scenario for MY 34 \citep{Montabone20}.

Except for the earliest profile measured at $\mathrm{L}_\mathrm{s}=163^\circ$, all profiles indicate that 
the O$_3$ density strongly increases at lower altitudes. This suggests a gradual transition towards the 
surface O$_3$ layer typically observed in the winter polar vortex. The 
largest O$_3$ densities reach 10$^{10}$~molec.\,cm$^{-3}$ at 10~km and below, in agreement with the polar profiles 
measured by the MEx SPectroscopie pour l'Investigation des Caract\'eristiques Atmosph\'eriques de Mars  (SPICAM) 
in the southern hemisphere \citep{Lebonnois06,Montmessin13}. 

SPICAM has also been used to measure the O$_3$ mixing ratio near the south polar enhancement 
\citep{Montmessin13,Piccialli19}. 
These latter authors observed strongly increasing O$_3$ abundance below 30~km toward 100--300~ppbv, which was distinct from 
mid-latitude observations. These mixing ratios are comparable to those presented in Fig.~\ref{fig:f5}b, which are 
between 100 and 300~ppbv. 

In Fig.~\ref{fig:f5}c, the LMD GCM ozone profiles co-located with the ACS measurements also show a 
large variability in the short period of $\mathrm{L}_\mathrm{s}$ sampled here. In general, the O$_3$ 
mixing ratios calculated by the model are underestimated by almost of factor of two relative to ACS MIR. 
This may reflect an imperfect timing in the model of the rapid decline in H$_2$O that accompanies 
the buildup of the northern polar vortex at this time of the year. Below 20~km, where ACS MIR observations 
are most sensitive, the model simulation shows a pronounced O$_3$ minimum  for the earlier profiles 
($\mathrm{L}_\mathrm{s}=163$--$189^\circ$) that is not seen in the measurements. 
The model outputs suggest that this problem is due to an overly strong poleward transport of H$_2$O-rich air 
originating from mid-latitudes. The observation presented in Fig.~\ref{fig:f5} in brown, 2476 N1, 
is an outlier in that ozone decreases with altitude above 10~km. The other occultations from this 
period feature an increasing or roughly constant mixing ratio up to 30~km. In the LMD GCM, the shape 
of the O$_3$ profile for this latest observation ($\mathrm{L}_\mathrm{s}=192.7^\circ$) is also 
characterised by a strong decrease above 20~km that contrasts with the earlier profiles. Examination 
of the model results shows that the change in the shape of this last O$_3$ profile is related to a large 
increase in H$_2$O above 20~km, which is also observed at the same time and location by the ACS NIR instrument 
\citep{Fedorova20}.

\section{Implications for CH$_4$ observation}
\label{sec:ch4}
Ozone absorption below 30~km in the mid-infrared range has important implications for searches for atmospheric 
methane. Past observations of methane in the atmosphere of Mars \citep{Formisano04,Kras04,Mumma09,Webster15} 
were a driving cause of the development of the ExoMars TGO mission. CH$_4$ should have 
a relatively short lifetime in the atmosphere of Mars (several hundred years), meaning current observations 
require an active source \citep{Lefevre09}. A key objective of the TGO mission is to determine with certainty 
whether or not CH$_4$ is present in the atmosphere of Mars and what its spatial and temporal variability is, and to 
localise any possible sources. This story continues to be intriguing as the first results from TGO reported 
an upper limit on the order of 50~pptv \citep{Korablev19}, and ACS MIR observations continue to reveal no 
methane after one MY. In its place, we have instead found the rare and previously undetected signatures
of O$_3$ and a new CO$_2$ magnetic dipole band \citep{Trokh20}.

\begin{figure*}[h!]
   \includegraphics[width=18cm]{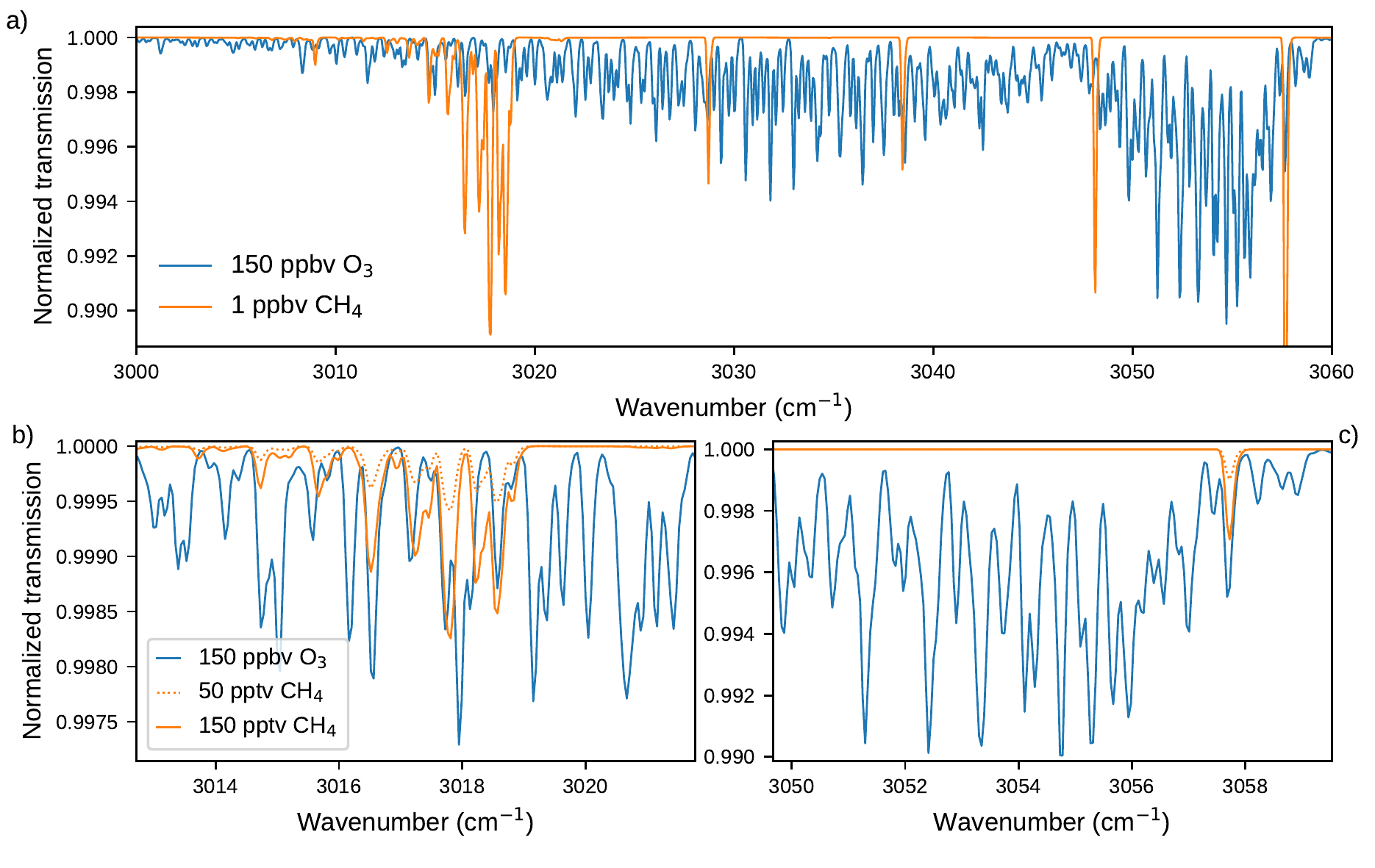}
   \caption{a) Modelled transmission spectrum contributions from 150~ppbv of O$_3$ and 1~ppbv of CH$_4$. 
   b) Close look at the modelled contributions around the CH$_4$ $Q$-branch, but using 50 and 150~pptv 
   of CH$_4$. c) Close look at the modelled contributions around the strongest line in the CH$_4$ $R$-branch.}
   \label{fig:f6}
\end{figure*}

The strongest observed O$_3$ features directly overlap important methane features used by the Planetary
Fourier Spectrometer (PFS) on Mars Express, the Tunable Laser Spectrometer (TLS) on Curiosity, and 
ground-based observatories. Figure~\ref{fig:f6}a compares the contributions to a transmission spectrum 
for 150~ppbv of O$_3$ and 1~ppbv of methane.  Figure~\ref{fig:f6}b shows a zoomed region surrounding 
the CH$_4$ $Q$-branch, and Fig.~\ref{fig:f6}c shows a zoomed region surrounding the strongest line of the 
CH$_4$ $R$-branch. Panels (a) and (b) also show 50--150~pptv of methane, approaching the upper limits reported 
by ACS \citep{Korablev19}. In both cases, there is direct overlap between absorption lines of CH$_4$ 
and O$_3$, with absorption lines produced by 150~ppbv O$_3$ being deeper than those from 150~pptv
of CH$_4$. Reported values of methane abundance range from the 
background measured by TLS of 0.4~ppbv \citep{Webster18} to enhancements measured by TLS and PFS of 
6--16~ppbv \citep{Giuranna19,Webster15}. We computed model spectra at the resolutions of both instruments 
and are confident that errors can be made when CO$_2$ and O$_3$ spectroscopy is 
not accounted for. 

The $Q$-branch is especially important for spectrometers with coarser spectral resolution than ACS MIR,
as the integrated CH$4$ lines that make up the $Q$-branch would have a higher magnitude than the 
individual lines in the $P$- and $R$-branches. 
It would be observed in order 180 of ACS MIR, shown in Fig.~\ref{fig:f1} and \ref{fig:f2}, and is 
located in the same spectral range in which we report previously unknown CO$_2$ lines \citep{Trokh20}, 
identified in Fig.~\ref{fig:f1}a. The $Q$-branch is used by the PFS team in their CH$_4$ analysis 
\citep{Formisano04,Giuranna19}, but with their spectral resolution, 1.3~cm$^{-1}$, the entire $Q$-branch
region is integrated into only two or three spectral points.

The $R$-branch consists of a series of broadly spaced lines, the strongest of which lies at 3057.7~cm$^{-1}$. This 
line has been used exclusively by TLS, which has sufficiently fine spectral resolution to observe the line as a 
triplet \citep{Webster15,Webster18}. The TLS instrument  also has the capability to resolve the overlapping 
O$_3$ line as a triplet. At this resolution, one pair of O$_3$ and CH$_4$ triplet lines directly 
overlap (3057.685~cm$^{-1}$), and the other two pairs partially overlap. Despite this, the TLS
team has not yet reported the abundance of ozone in Gale crater.

CO$_2$ and O$_3$ alone cannot account for the detections made by both teams. In the case of PFS, the 
previously unknown CO$_2$ features would impact all observations equally, as CO$_2$ is always present and 
well-mixed. The PFS team has instead identified CH$_4$ in only a small number of observations 
\citep{Formisano04,Giuranna19}. Furthermore, 
we computed spectra with O$_3$ at two and three times the quantities in our observations, and the 
sheer magnitude of CH$_4$  observed by these latter authors (15~ppbv) is far too large to be easily mistaken for O$_3$.

In the case of TLS, which takes measurements of CH$_4$ at the surface and mostly at night where and when 
the O$_3$ abundance is greatest, again, it is unlikely that large quantity of CH$_4$ observed (up to 9~ppbv)
resulted from O$_3$, yet the latter may interfere in the measurement of the background 
level of methane in the so-called enriched mode as both ozone and methane should sustain the same enrichment.

For ground-based observations, strong O$_3$ absorption features from Earth's atmosphere must first be  removed 
before retrieving mixing ratios for Mars \citep{Kras12,Mumma09};  O$_3$ must be accounted for,
although this step makes the retrieval more difficult \citep{Zahnle11}.
Finally, in the case of  all previous observations, the rapid evolution 
and disappearance of CH$_4$ are still not explained, although ozone chemistry is very rapid, with a lifetime 
on the order of days.

\section{Conclusion}
The faint spectral signature of ozone, an established trace gas in the Martian atmosphere, has been observed 
for the first time in the MIR spectral region by the ACS MIR instrument on ExoMars TGO. These 
observations are limited to high northern latitudes ($>65^\circ$N) and prior to the onset of the 2018 Mars 
global dust storm. During this time period, ACS MIR measurements provide new insight into the vertical structure 
of ozone around the northern fall equinox and show the variability its VMR can have shortly before the polar 
vortex is established. We observe the distinct presence of ozone with 100--200~ppbv at 20~km and below, which is close 
to the amounts measured in comparable conditions at the edge of the (southern) polar vortex. 
In general, the O$_3$ mixing ratios retrieved by ACS MIR are higher than those calculated by the LMD GCM. 
This seems to result from an overly wet atmosphere in the model at the time (equinox) and location 
(high northern latitudes) sampled here; these parameters will need to be confirmed in the future by simultaneous measurements of 
water vapour by ACS NIR.

We look forward to the processing of observations made in MY 35 and southern winter. Dust obscured the northern 
ACS observations after $\mathrm{L}_\mathrm{s}190^\circ$, but observations made after 
$\mathrm{L}_\mathrm{s}\sim30^\circ$ in the next MY are very clear, and we expect to be able to detect 
low-altitude ozone near the southern hemisphere.

The observation of this trace gas at higher-than-predicted VMRs   below 30~km has important implications for the 
detection of methane in the atmosphere of Mars. This band, as well as the previously unidentified CO$_2$ band, 
overlap and interfere with the CH$_4$ $\nu_3$ band used by TGO, MEx, and MSL to search for methane. Accounting 
for these absorption features improves our own spectral fitting and will lead to more accurate lower limits 
in the future. In conclusion, our study shows that the O$_3$ lines we report here interfere with measurements 
of Martian methane, but a detailed reanalysis of these measurements is required to precisely assess their impact.

\begin{acknowledgements}
   The ACS investigation was developed by the Space Research Institute (IKI) in Moscow, and the Laboratoire 
   Atmosph\`eres, Milieux, Observations Spatiales (LATMOS/CNRS) in Paris. The investigation was funded by 
   Roscosmos, the National Centre for Space Studies of France (CNES) and the Ministry of Science and Education 
   of Russia. The GGG software suite is maintained at JPL (tccon-wiki.caltech.edu). This work was funded by 
   the Natural Sciences and Engineering Research Council of Canada (NSERC) (PDF--516895--2018), the UK Space 
   Agency (ST/T002069/1) and the National Centre for Space Studies of France (CNES).

   All spectral fitting was performed by KSO using the GGG software suite. The interpretation of the results was 
   done by KSO and FL. The processing of ACS spectra is done at IKI by AT and at LATMOS by LB. Input and aid on 
   spectral fitting were given by JA, DB, AF, AL, and FM. The ACS instrument was designed, developed, and 
   operated by AP, AS, AT, FM, and OK.
\end{acknowledgements}

\bibliographystyle{aa}

\end{document}